\documentclass[preprint2]{aastex}

\begin{document}

\title{A multi-resolution, multi-epoch low radio frequency survey of the Kepler K2 mission Campaign 1 field}

\author{S.J. Tingay\altaffilmark{1,2},  P.J. Hancock\altaffilmark{1,2}, R.B. Wayth\altaffilmark{1,2}, H. Intema\altaffilmark{3}, P. Jagannathan\altaffilmark{4,5}, K. Mooley\altaffilmark{6}}

\altaffiltext{1}{International Centre for Radio Astronomy Research (ICRAR), Curtin University, Bentley, WA 6102, Australia}
\altaffiltext{2}{ARC Centre of Excellence for All-sky Astrophysics (CAASTRO), Sydney, Australia} 
\altaffiltext{3}{Leiden Observatory, Universiteit Leiden, Leiden, The Netherlands} 
\altaffiltext{4}{National Radio Astronomy Observatory, 1003 Lopezville Road, Socorro, NM 87801-0387, USA }
\altaffiltext{5}{Department of Astronomy, University of Cape Town, Rondebosch, Cape Town, 7700, South Africa}
\altaffiltext{6}{Centre for Astrophysical Surveys, University of Oxford, Denys Wilkinson Building, Keble Road, Oxford, OX1 3RH}


\begin{abstract}
We present the first dedicated radio continuum survey of a Kepler K2 mission field, Field 1 covering the North Galactic Cap.  The survey is wide field, contemporaneous, multi-epoch, and multi-resolution in nature and was conducted at low radio frequencies between 140 and 200 MHz.  The multi-epoch and ultra wide field (but relatively low resolution) part of the survey was provided by 15 nights of observation with the Murchison Widefield Array (MWA) over a period of approximately a month, contemporaneous with K2 observations of the field.  The multi-resolution aspect of the survey was provided by the low resolution (4\arcmin) MWA imaging, complemented by non-contemporaneous but much higher resolution (20\arcsec) observations using the Giant Metrewave Radio Telescope (GMRT).  The survey is therefore sensitive to the details of radio structures across a wide range of angular scales.  Consistent with other recent low radio frequency surveys, no significant radio transients or variables were detected in the survey.  The resulting source catalogs consist of 1,085 and 1,468 detections in the two MWA observation bands (centered at 154 and 185 MHz, respectively) and 7,445 detections in the GMRT observation band (centered at 148 MHz), over 314 square degrees.  The survey is presented as a significant resource for multi-wavelength investigations of the more than 21,000 target objects in the K2 field.  We briefly examine our survey data against K2 target lists for dwarf star types (stellar types M and L) that have been known to produce radio flares.
\end{abstract} 


\section{INTRODUCTION}
The highly successful Kepler spacecraft is being utilised for the so-called K2 mission, a survey of a number of fields lying along the ecliptic that are accessible to the spacecraft with a reduced number of reaction wheels \citep{how14}; more than 10 observing campaigns are expected\footnote{http://keplerscience.arc.nasa.gov/K2/Fields.shtml}.

The K2 mission became operational in mid-2014, reaching a photometric precision approaching that obtained by the Kepler mission.  Each of the K2 observing campaigns cover an approximate 80 day period, with target lists compiled from community solicitations for proposed targets.

In this paper we describe contemporaneous observations of K2 Field 1 with the Murchison Widefield Array (MWA: \citet{tin13, lon09}) and historical (from 2010-12) observations from the Giant Metrewave Radio Telescope (GMRT) TIFR GMRT Sky Survey (TGSS\footnote{http://tgss.ncra.tifr.res.in/}), via the TGSS Alternative Data Release 1 (ADR1) \citep{int16}.  The MWA and GMRT are radio telescopes operating at low radio frequencies (approximately 140 - 200 MHz for the work described here), the MWA being the low frequency precursor for the Square Kilometre Array (SKA: \citet{dew09}).  

We describe multi-epoch observations with the MWA.  The MWA's extreme field of view encompasses the entirety of the K2 field, allowing efficient monitoring of all radio sources within Field 1.  We also describe historical, but significantly higher angular resolution, GMRT TGSS observations.  The combination of MWA and GMRT data provide a multi-epoch (via the MWA) and multi-resolution (low resolution MWA and high resolution GMRT) low radio frequency catalog for K2 Field 1.  The MWA observations were also coordinated with SkyMapper observations of K2 Field 1.  The SkyMapper observations will be published elsewhere (Wolf et al., in preparation).

A range of mechanisms will produce variability and transient phenomena at low radio frequencies, relevant to several areas of K2 science.  For example, flares from dwarf stars \citep{ber06,jae11} are expected to produce low frequency radio emission.  Low frequency flares similar to those observed from Jupiter may be expected from exoplanets \citep{zar01,mur15}.  And on longer timescales, active galactic nuclei are variable across the entire electromagnetic spectrum, including at low radio frequencies.  Overviews of classes of variable and transient radio sources relevant to K2 science can be found in \citet{Murphy_vast:2013} and \citet{bow13} (particularly for the low frequencies of the MWA and GMRT).

K2 mission Campaign 1 was conducted on Field 1 (center at RA=11:35:45.51; DEC=+01:25:02.28; J2000), covering the North Galactic Cap, between 2014 May 30 and 2014 August 21.  A target list of 21,649 targets across 76 proposals was constructed.  The target list was drawn from proposals seeking to study a wide range of galactic and extragalactic populations.  Reduced light curves for K2 Campaign 1 are available for the target list from https://www.cfa.harvard.edu/\~avanderb/k2.html, according to the analysis of \citet{Van14}.

In the future, the release of the MWA GLEAM continuum survey, covering the full observable sky from the MWA, will greatly expand the scope of the type of multi-wavelength studies described here \citep{way15}.  In particular, coupled with the GMRT TGSS, detailed multi-resolution low frequency cross matching will be possible over a large fraction of the sky.  Further, the LOFAR MSSS survey will play a similar role in the Northern Hemisphere \citep{hea15}.

The radio catalog is presented to support a range of multi-wavelength studies of objects within K2 Field 1.

\section{OBSERVATIONS}

\subsection{MWA}
The parameters of the MWA observations are described in Table 1, 15 observations conducted over a period of approximately one month in June and July 2014.  All observations were made in a standard MWA imaging mode with a 30.72 MHz bandwidth consisting of 24 contiguous 1.28 MHz ``coarse channels", each divided into 32 ``fine channels" each of 40 kHz bandwidth (total of 768 fine channels across 30.72 MHz).  The temporal resolution of the MWA correlator output was set to 0.5 s.  All observations were made in full polarimetric mode, with all Stokes parameters formed from the orthogonal linearly polarised feeds.

Observations were made at two centre frequencies, 154.88 MHz and 185.60 MHz, with two 296 s observations of the K2 field at each frequency on each night of observation, accompanied by observations of one of three calibrators (Centaurus A, Virgo A, or Hydra A) at each frequency, with 112 s observations.

The observed fields were tracked and thus, due to the fixed delay settings available to point the MWA primary beam, the tracked right ascension and declination changes slightly between different observations (always a very small change compared to the MWA field of view).

The total volume of MWA visibility data processed was approximately 2.2 TB.

\begin{table*}[ht]
  \begin{tabular}{c c c l c r r} \hline 
OBSID & START DATE/TIME (UT) & T (s) & TARGET & FREQ (MHz) & RA ($^{\circ}$) & DEC ($^{\circ}$)\\ \hline
1090489536 & 2014-07-27 09:45:20 & 112 & CenA & 185.60 & 198.87 & -40.09\\ 
1090489416 & 2014-07-27 09:43:20 & 112 & CenA & 154.88 & 198.37 & -40.08\\ 
1090489112 & 2014-07-27 09:38:16 & 296 & K2 & 185.60 & 169.69 & -0.31\\ 
1090488816 & 2014-07-27 09:33:20 & 296 & K2 & 154.88 & 168.46 & -0.31\\ 
1090488512 & 2014-07-27 09:28:16 & 296 & K2 & 185.60 & 175.10 & 5.26\\ 
1090488216 & 2014-07-27 09:23:20 & 296 & K2 & 154.88 & 173.87 & 5.26\\ \hline
 \end{tabular}
   \caption{MWA observation log: Column 1 - MWA observation ID; Column 2 - UT date and time of observation start; Column 3 - duration of observation in seconds; Column 4 - observation target (K2=K2 Field 1; CenA = Centaurus A; HydA=Hydra A); Column 5 - centre frequency in MHz; Column 6 - Right Ascension in decimal degrees; and Column 7 - Declination in decimal degrees.  Only the entries for one of the 15 nights of observation is shown here.  The full observation table is available as a Machine Readable Table (CSV format).}
\end{table*}

\subsection{GMRT TGSS}
A full survey of the radio sky at 150 MHz as visible from the GMRT was performed within the scope of the PI-driven TGSS project between 2010 and early 2012, covering the declination range $-$55 to $+$90 degrees. Summarizing the observational parameters as given on the TGSS project website\footnote{http://tgss.ncra.tifr.res.in/150MHz/obsstrategy.html}, the survey consists of more than 5000 pointings on an approximate hexagonal grid. Data were recorded in full polarization (RR,LL,RL,LR) every 2 seconds, in 256 frequency channels across 16 MHz of bandwidth (140--156 MHz). Each pointing was observed for about 15 minutes, split over three or more scans spaced in time to improve UV-coverage. Typically, 20--40 pointings were grouped together into single night-time observing sessions, bracketed and interleaved by primary (flux density and bandpass) calibrator scans on 3C48, 3C147 and/or 3C286. Interleaving secondary (phase) calibrator scans on a variety of standard phase calibrators were also included, but were typically too faint to be of significant benefit at these frequencies.

To date, in TGSS data release 5, only about 10~percent of the survey images have been processed and released through the TGSS project website\footnote{http://tgss.ncra.tifr.res.in/150MHz/download.html}. The slow progress reflects the difficulty of processing high-resolution, low-frequency radio observations without taking into account direction-dependent effects (DDEs) like ionospheric phase delay. Given the importance of TGSS as a reference catalog for aperture array telescopes like MWA and the LOw Frequency ARray \citep[LOFAR;][]{van14}, and the fact that all survey data is well outside the proprietary period and freely available in the GMRT archive, \citet{int16} have independently (re-)processed the full survey, including direction-dependent calibration and imaging (see Section~\ref{sec:tgss_data_analysis}). The resulting images used in this study are part of this effort.

\section{DATA ANALYSIS AND RESULTS}

\subsection{MWA DATA ANALYSIS}
The raw visibility data were pre-processed through the `cotter' pipeline \citep{2015PASA...32....8O} which performs automated flagging of any radio-frequency interference (RFI) using the `AO Flagger' \citep{2012MNRAS.422..563O} software and converts the data to the standard uvfits format.  Subsequent processing was performed with the \textsc{miriad} \citep{1995ASPC...77..433S} data processing package (Version 1.5, recompiled in order to allow 128 antenna arrays).

The datasets in Table 1 were processed in an imaging and calibration pipeline implemented using \textsc{Miriad} task calls from a PERL script.

\subsubsection{Initial calibration}
The calibrator and target data were read into \textsc{miriad} using task \textsc{fits} and the task \textsc{uvaver} was used to split the XX and YY polarisations into separate datasets for calibration and imaging.  The calibrator name was identified from the FITS file header and used to determine initial calibration steps.  

If the calibrator was identified as Hydra A or Virgo A, task \textsc{selfcal} was used to calculate antenna-based gains for each of the 24$\times$1.28 MHz coarse channels (each polarisation separately).  At the MWA resolution, a point source model for both Hydra A and Virgo A is adequate to obtain reasonable calibration solutions.  The point source flux density used for Hydra A was parameterised as $328.7\frac{\nu}{150 MHz}^{-0.91}$ Jy, where $\nu$ is the observing frequency in MHz.  Similarly, for Virgo A, the point source flux density was parameterised as $1450\frac{\nu}{150 MHz}^{-0.86}$ Jy.  Flux density models for both Hydra A and Virgo A are from \citet{bar77}; although the \citet{bar77} scale does not extend below 350 MHz, recent work confirms that an extrapolation to 150 MHz is valid (Perley et al. 2016, in preparation).  \textsc{selfcal} was run with a five minute solution interval (greater than scan length), calculating amplitude and phase components of the gain correction for each of the 24$\times$1.28 MHz coarse channels in the visibility data.

The \textsc{selfcal} solutions were applied to the data using task \textsc{uvaver} and the data were imaged using task \textsc{invert} (4096$\times$4096 pixel images with 30\arcsec~ pixels, robust weighting of 0.5, multi-frequency synthesis turned on, and producing a dirty beam double the size of the image).  These images were cleaned with task \textsc{clean} using a speed of -1 and 10,000 iterations.  The resulting clean models were then used in \textsc{selfcal}, using the same options as above, to refine the calibration solutions.

These calibration solutions were transferred to the corresponding target datasets and applied to the target visibilities using \textsc{uvaver}.

If the calibrator was determined to be Centaurus A, a point source model was inadequate due to the highly extended nature of the calibrator.  In this case, a model of the Centaurus A brightness distribution \citep{cena} was scaled to the MWA frequency according to $\frac{\nu}{843 MHz}^{-0.50}$, again where $\nu$ is the MWA observing frequency in MHz.  This model was used in \textsc{selfcal}, with the same options as listed above, but with an additional restriction that only visibilities on projected baselines longer than 0.05 k$\lambda$ were used to find self-calibration solutions.

It was found that no further self-calibration was required in this case.  Thus, the self-calibration solutions were transferred to the corresponding target dates and applied, as above, using \textsc{uvaver}.

\subsubsection{Imaging of target fields}
The calibrated target data were imaged using task \textsc{invert}, with image sizes of 7000$\times$7000 pixels, pixel sizes of 30\arcsec (resulting in images of approximately $70^{\circ}\times70^{\circ}$ = 4900 sq. deg.), robust weighting of 0.5, with the multi-frequency synthesis option turned on, dirty beams twice the images produced, and with the spectral dirty beam production option turned on.  Production of the spectral dirty beam allows the use of \textsc{mfclean}, a version of clean in \textsc{miriad} that returns the spectral index as well as the brightness distribution of objects cleaned.  \textsc{mfclean} was run with a speed of -1, 10,000 iterations, and only cleaning the inner 66\% of the images (hence the choice of the large image sizes in \textsc{invert}).

A powerful radio galaxy, M87, lies near the edge of the MWA field of view for the target pointing direction, producing significant sidelobes across the imaged field of view for snapshot observations of this type.  In order to minimise the effect of M87 the task \textsc{mfclean} was used (described above) in an attempt to solve for spectral effects due to the primary beam shape in the direction of M87.  This was largely successful, but locally around M87, sidelobes were still being cleaned and included in the clean component models.

In an attempt to further reduce the effect of M87 and the attendant sidelobes, a small area around M87 was defined from which false clean components were excised from the clean component model.  \textsc{selfcal} was run on the target dataset using the modified clean component model, solving for both amplitude and phase.  The resultant dataset was then reimaged using \textsc{mfclean} as described above.  This provided a small but noticeable improvement in the final images.

All images were then restored using task \textsc{restor} with default options.  Keywords were then inserted into the target data headers to properly describe the slant orthographic projection required for the MWA field of view, using task \textsc{puthd}.  The restored and header-corrected images were written to disk as FITS images.  

A script that calculates the MWA primary beam pattern from the FITS images was used to calculate a primary beam corresponding to every target observation \citep{sut15}.  The beams were also stored as FITS images.  The beams were read into \textsc{miriad} using task \textsc{fits} and the task \textsc{math} was used to correct the images derived from \textsc{mfclean} and \textsc{restor} with their corresponding primary beams (different beams are produced and applied for each of the XX and YY images).

In all primary beam corrected images, the flux density of a known radio source was measured, 3C 257.  The flux density of 3C 257 was modelled as $886.6\nu^{-0.89}$ Jy (with $\nu$ in MHz), derived from available flux density measurements from the literature between 38 MHz and 750 MHz \citep{pau66,kel69,lar81,sle95,dou96,coh07,jac11}.  A final correction to the flux density scale was derived from a comparison of the calculated and measured flux densities.  The corrections were applied as a scaling to the images using task \textsc{math}.  Finally, the task \textsc{math} was used to combine XX and YY images into Stokes I images, which were then read out to disk for source finding and further analysis.

The synthesised beam at 154 MHz is approximately 4.6\arcmin $\times$ 4.2\arcmin~ at a position angle of 105$^{\circ}$ and approximately 4\arcmin$\times$3\arcmin~ at a position angle of 109$^{\circ}$ at 185 MHz.  The 154 MHz images have a typical noise of 100~mJy/beam, whilst the 184 MHz images have a typical noise of 70~mJy/beam.

An example image for a single snapshot observation at 154 MHz is shown in Figure 1.

\begin{figure*}[ht]
\centering
\includegraphics[width=0.95\linewidth,angle=0]{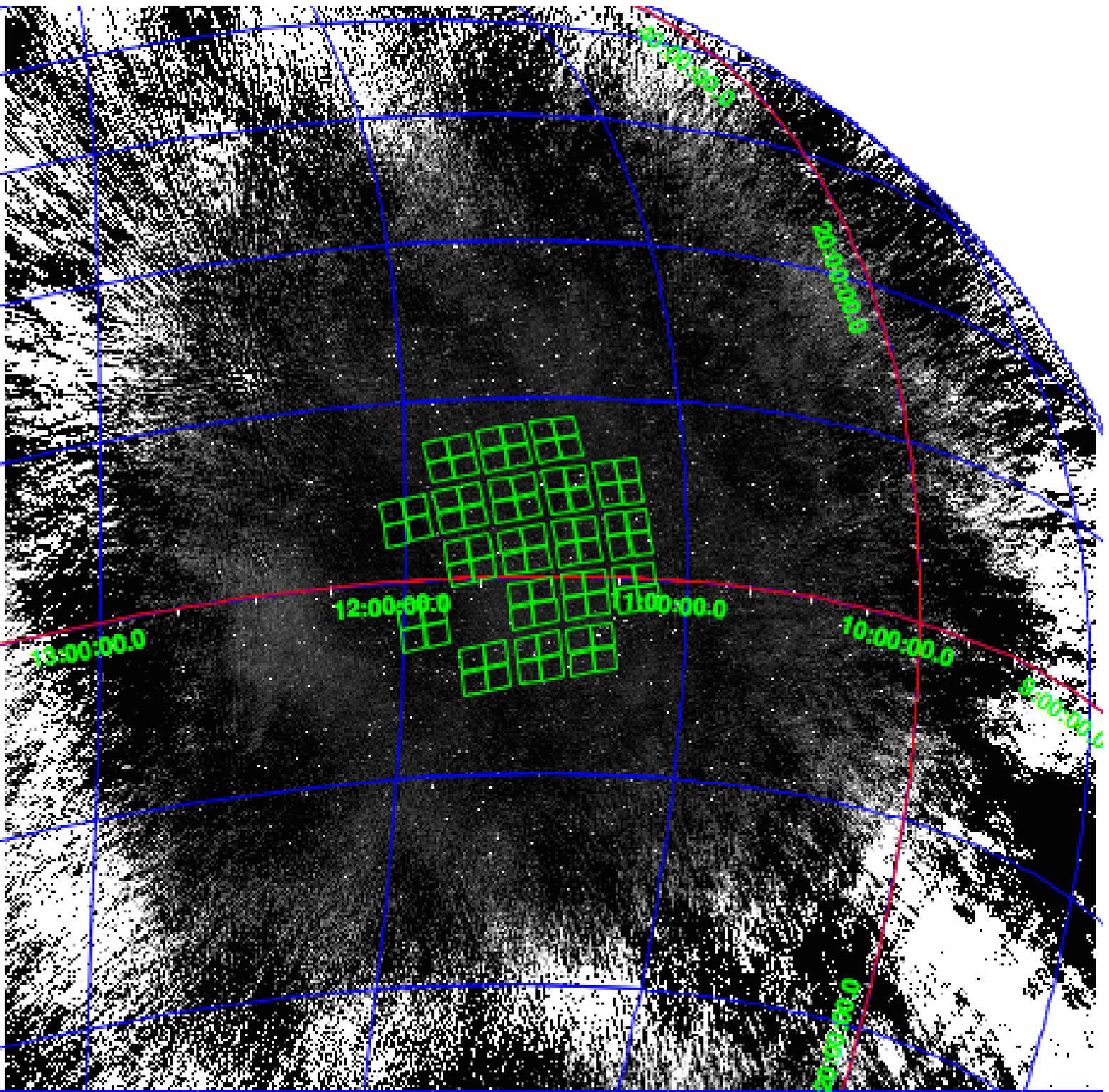}
\caption{An example snapshot Stokes I image of the MWA field, at a centre frequency of 154 MHz.  The synthesised beam is approximately 4.6\arcmin $\times$ 4.2\arcmin at a position angle of 105$^{\circ}$.  M87 can be seen near the top left of the primary field of view.  The greyscale levels are set to range between -0.2 Jy/beam and 2 Jy/beam.  The noise level near the middle of the field of view is approximately 100 mJy.  The footprint of the K2 mission field of view for Campaign 1 is shown, overlaid.}
\end{figure*}

\subsubsection{Source finding and production of light curves}
\label{sec:source_finding}

The production of light curves proceeded in four main stages: characterizing
the background and noise properties of each image; source finding within each
image; cross-matching sources across multiple images; and correcting the flux density scale
between different pointing directions. Each step is described below.

The background and noise properties of each image were characterized using the
Background and Noise Estimation program (BANE\footnote{Available at \url{GitHub.com/PaulHancock/Aegean}}). Figure 2 shows the resulting background and noise properties for just one of the MWA images. The noise image is able to account for cleaning artifacts around bright sources in our error budget, as well as the larger scale variance in the image. The background image shows some large scale structures. These structures are not imaging or observing artifacts, but in fact real extended emission. The features that are seen in Figure 2 are found to correlate strongly with features present in the point source subtracted Haslam all sky 408 MHz map \citep{has82}.

BANE characterises both the background and noise properties of an image, as the median and standard deviation of the image pixels, respectively. The median and standard deviation are calculated over a box that is 30 times the size the synthesized beam on a side. Sigma clipping is used to remove the contribution of sources from the pixel distribution before the median and standard deviation are calculated. The source characterisation that is performed by Aegean, is done on a background subtracted image.

The 154 MHz images have a typical noise of 100~mJy/beam, whilst the 184 MHz images have a typical noise of 70~mJy/beam. The increased sensitivity of the 185 MHz images means that more sources are found in these images, and thus not all sources have counterparts at both frequencies.

After the background and noise images were created, we used the Variable and Slow Transients \citep[VAST,][]{Murphy_vast:2013} pipeline \citep{Banyer_vast:2012}, to coordinate the source finding and cross matching. The source finding stage uses the Aegean source finding algorithm \citep{hancock_compact_2012} version 1.9.6\footnote{Available at \url{GitHub.com/PaulHancock/Aegean}}. The cross matching of sources between epochs and datasets is done using a Bayesian approach that is described by \citet{budavari_cross-identification_2011}.  We used a 3$\sigma$ detection threshold.

The MWA observations were conducted at a range of different pointing
directions on a given night. Though the pointing directions change between
observations within a single night, and between different nights, each
pointing direction is observed at the same LST. Not all pointing directions
currently have a well established primary beam model \citep{sut15}. Thus the
relative flux density calibration within a single pointing center is
reproducible, however the calibration between different pointing centers is
not. Thus changing pointing centers within a night of observations can
introduce systematic discontinuities in all of the light curves. To achieve a
self consistent flux density scale between pointing directions we grouped
observations into sets that shared the same pointing center. Then for each
source we computed the mean flux density within each group of observations,
and adjusted the flux densities so that each group of observations had the
same mean flux density. In this way we were able to remove the discontinuities
that were introduced by the changing pointing centers, whilst at the same time
creating a single light curve for each source that could be analyzed for
variability. In Figure 3 we show light curves for four sources: the primary
flux density calibrator ($4C+05.49$); two bright 4C sources in different parts
of the image (4C$-$0.42 and 4C$+$04.37); and a faint source that is near the
flux density calibrator, but not within the 4C catalog. In each case, it can
be seen that the variability due to changing pointing centers has been
removed.

Figure 4 shows the mean normalised light curve and 1$\sigma$ bounds at 154 MHz, over all epochs and pointing directions, showing that the flux density calibration has not introduced variability into the light curves.

A different primary beam model is required for each of the MWA frequencies,
each of which suffer from the effects mentioned in the previous paragraph. The
flux densities at each frequency have been adjusted so that the primary flux
density calibrator is correct at each frequency. However, since the primary beam model
is uncertain by different amounts over the field of view this means that the
flux densities of non-calibrator sources are not generally consistent across frequencies; the
spectral indices of sources between the two frequencies become more uncertain
the further the source is from the flux density calibrator.  We therefore warn users of the catalogs to exercise care in using spectral indices calculated from the flux densities in the catalogs.

Figure 5 shows the measured spectral index variations as a function of position along a south-east/north-west line through the image.  Detected sources toward the south-east have, on average, flatter spectral indices than toward the north-west.  No attempt has been made to correct for this effect in the catalogs, as it is not a primary consideration for the catalogs and better spectral index measurements will be available via the GLEAM survey, described in \cite{way15} and Hurley-Walker et al. (2016, in preparation).

We also examined coordinate offsets in the MWA images and found no significant deviations from zero offset as a function of position in the images.  Using the GMRT positions as references, we found a mean position offset of 3\arcsec$\pm$50\arcsec~ in right ascension and 5\arcsec$\pm$60\arcsec~ in declination at 154 MHz, and -3\arcsec$\pm$40\arcsec~ in right ascension and 3\arcsec$\pm$50\arcsec~ in declination at 185 MHz.

\begin{figure*}[ht]
\centering
\includegraphics[width=0.95\linewidth,angle=0]{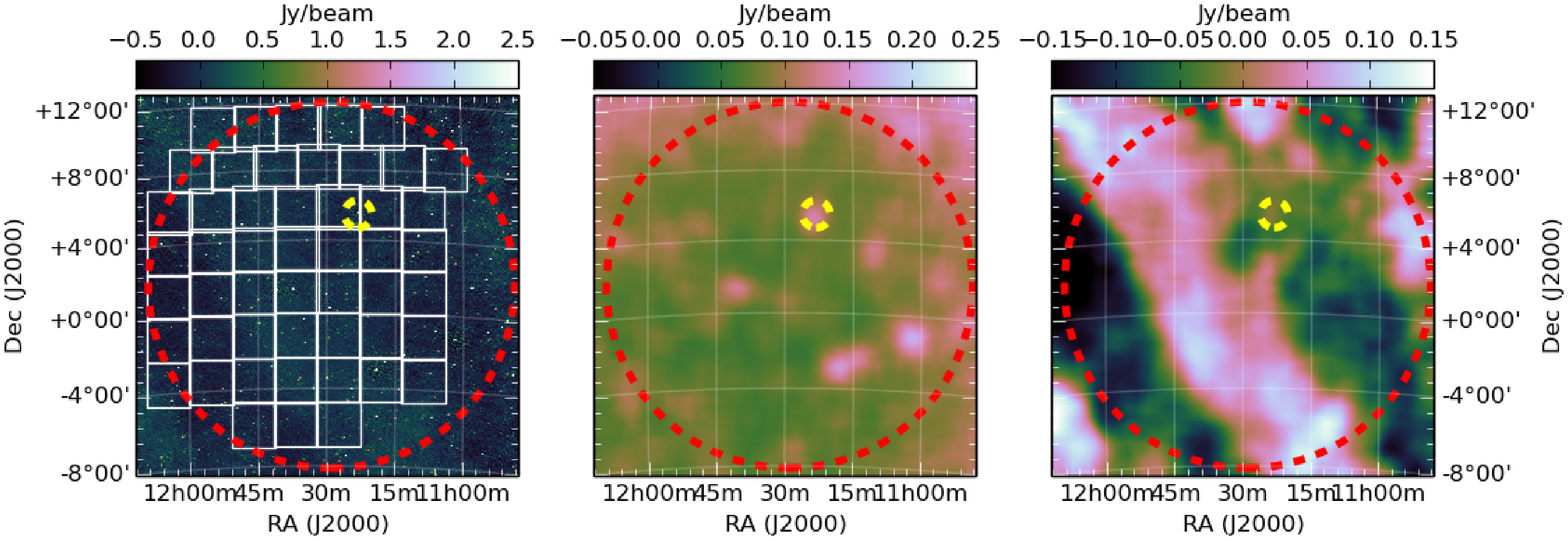}
\caption{Three images of the region of interest. The red circle is the region from which we select sources from for source finding. The yellow circle is the location of the flux density calibrator. The left panel is the sky image, with SkyMapper observing fields overlaid. The middle panel is the noise image, and the right panel is the background image. All the flux density scales are different. The noise enhancements in the noise image are due to side-lobes around bright sources.  The emission that is seen in the background image correlates with features in the Haslam 408 MHz all sky map.}
\end{figure*}

\begin{figure*}[ht]
\centering
\includegraphics[width=12cm,angle=0]{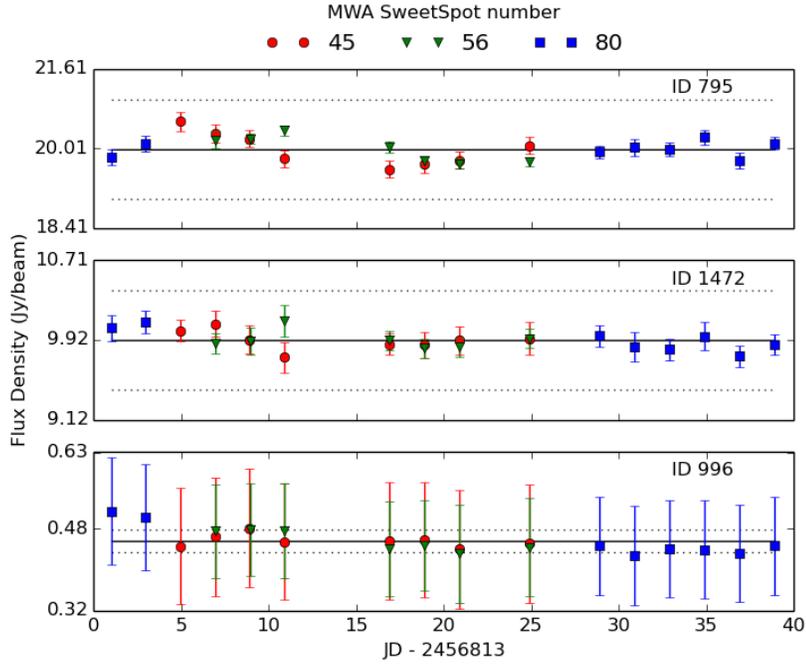}
\caption{The light curves of three sources from the 154MHz catalogue. Error bars are the $1\sigma$ measurement errors. Observations from different telescope pointings (sweet spots) are identified by colour/shape. The horizontal lines show the median flux of the source across all epochs (solid) and a 5\% variation about this flux (dotted). Source 795 is the brightest source in the catalogue, source 1472 is located at the edge of the region of interest, and source 996 is a faint source near the centre of the image. The light curve of all sources is clearly affected by pointing changes, however no variability is seen beyond the 5\% level.}
\end{figure*}

\begin{figure*}[ht]
\centering
\includegraphics[width=12cm,angle=0]{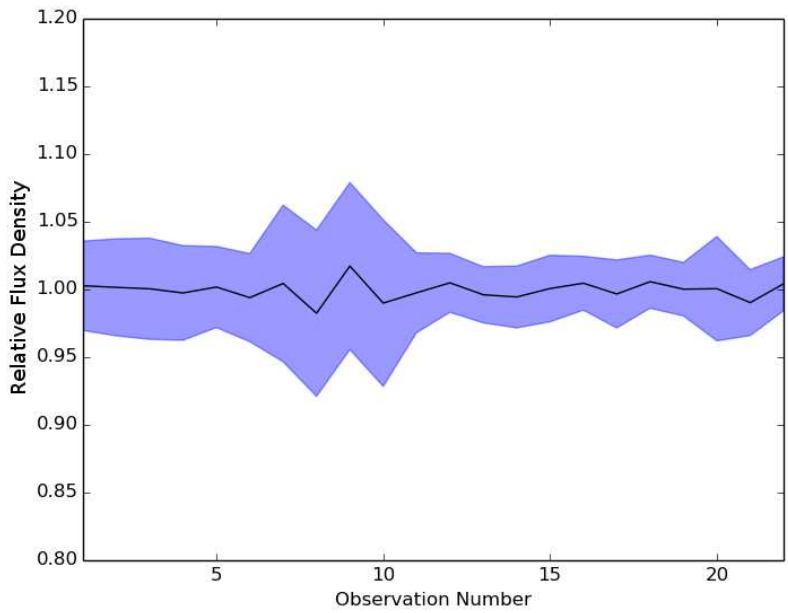}
\caption{The mean light curve of all sources at 154~MHz over all epochs and pointing directions as a function of observation number. The black line is the mean of the normalized light curves, and the blue region represents a $1\sigma$ deviation. The average light curve is thus consistent with no variability to a level of $\sim5\%$, confirming that the flux density calibration method has not introduced variability despite the different pointing directions used.}
\end{figure*}

\begin{figure*}[ht]
\centering
\includegraphics[width=12cm,angle=0]{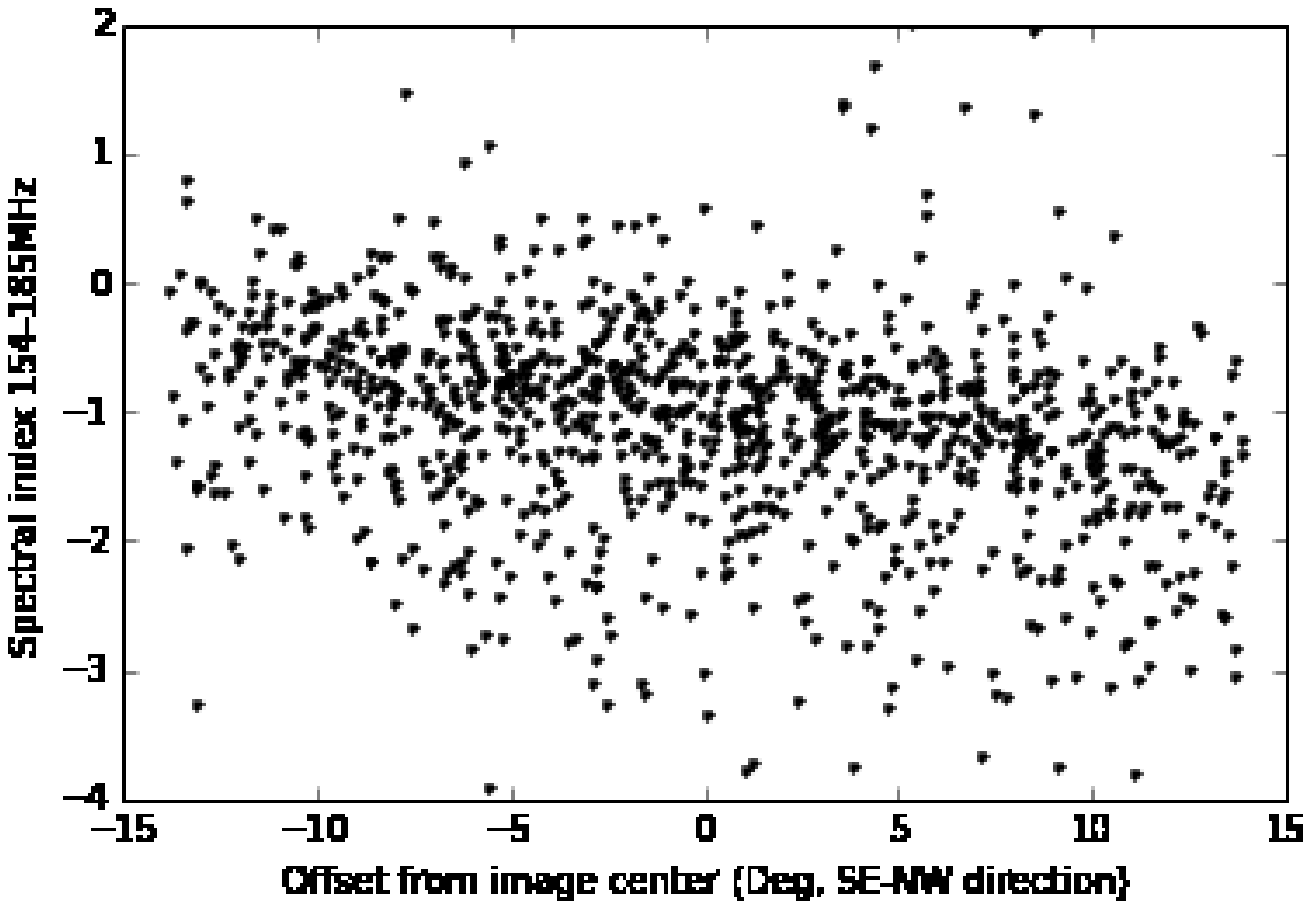}
\caption{The spectral index variations as a function of position in the field of view, induced by residual errors in the primary beam model far from the pointing centre.}
\end{figure*}

\subsection{TGSS DATA ANALYSIS}
\label{sec:tgss_data_analysis}
Archival TGSS data were obtained from the GMRT archive, and processed with a fully automated pipeline based on the SPAM package \citep{int09,int14}, which includes direction-dependent calibration, modeling, and imaging of ionospheric phase delay. In summary, the pipeline consists of two parts: a \emph{pre-processing} part that converts the raw data from individual observing sessions into pre-calibrated visibility data sets for all observed pointings; and a \emph{main pipeline} part that converts pre-calibrated visibility data per pointing into stokes I continuum images. Both parts run as independent processes on multi-node, multi-core compute clusters, allowing for significant parallel processing of many observations and pointings at the same time. Making use of the compute cluster at the National Radio Astronomy Observatory (NRAO) in Socorro NM (USA), processing of all TGSS survey data was completed well within a month. TGSS image mosaics (like the MWA K2 field) have been made available upon request\footnote{Contact \email{intema@strw.leidenuniv.nl}}, mainly because web-based access to the data products is not yet in place. More details on the processing pipeline and characteristics of the survey are given in \citep{int16}.

The single epoch TGSS image was processed in the same way as each of the MWA images using the background and noise characterization source finding techniques outlined in \ref{sec:source_finding}.

Figure 6 shows a small portion of the GMRT image, within the MWA field of view.

\begin{figure*}[ht]
\centering
\includegraphics[width=0.95\linewidth,angle=0]{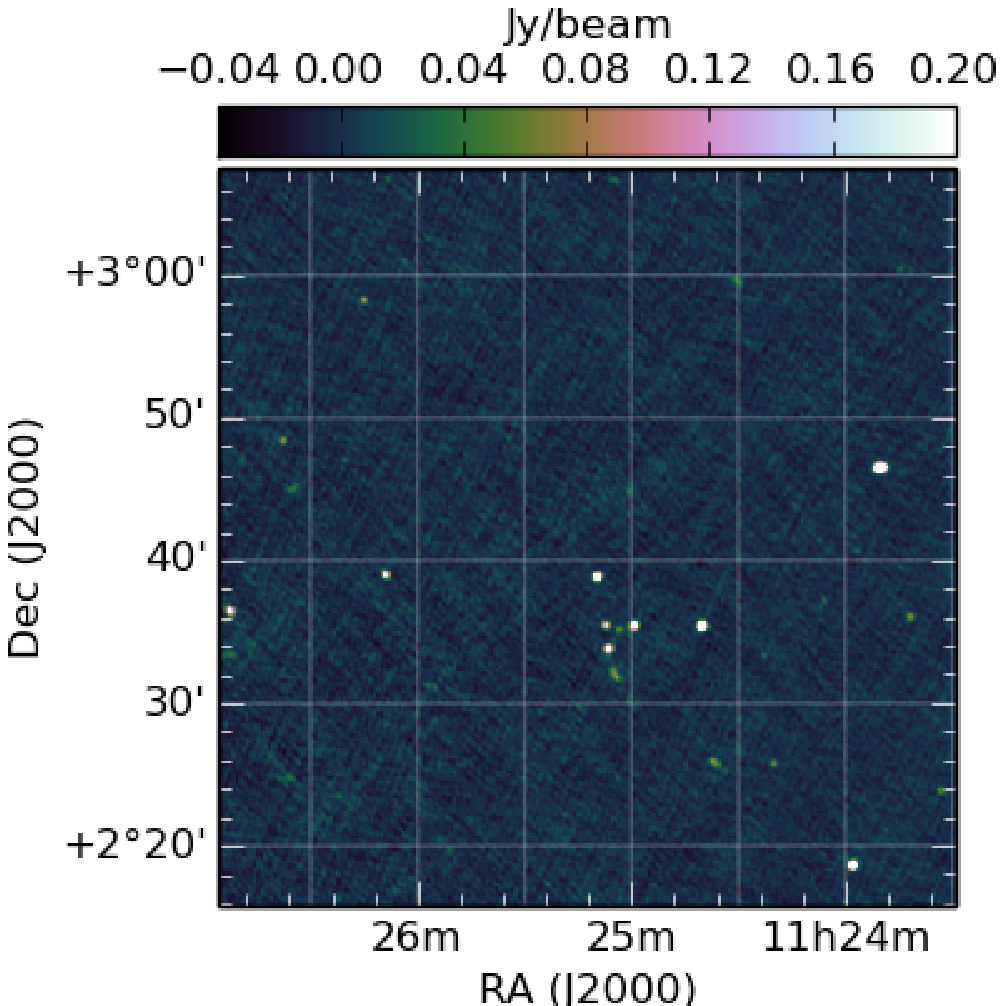}
\caption{A small portion of the GMRT image, within the MWA field of view.  The synthesised beam is circular with a FWHM at 21\arcsec.  The noise level is approximately 7 mJy/beam.}
\end{figure*}

\section{DISCUSSION AND CONCLUSION}
\label{sec:catalog}

A source catalog was produced from each of the two frequencies of MWA data and
the single TGSS image. For the MWA data, the source flux densities were averaged over
all epochs to produce a single flux density measurement. Since the 154~MHz and 185 MHz
images have different sensitivities, and slightly different resolutions,
separate catalogs were created for the MWA data. There are no extended or
resolved sources in the MWA catalogs so only the flux density is reported (equal to the peak brightness in the image for a point source). For
the higher resolution TGSS images, sources were resolved in some cases and so
morphology information is included in this catalog. Thus, there are three catalogs
in total. In Table~\ref{tab:cat_154} we show a portion of the MWA catalog at
154~MHz for reference (the 185 MHz catalog has the same format), and in Table~\ref{tab:cat_TGSS} we show a portion of the catalog based on the GMRT TGSS images.  All three catalogs are provided as Machine Readable Tables (in CSV format).  In addition, we provide two further Machine Readable Tables (also in CSV format) that contain the light curve information for the MWA catalogs at 154 and 185 MHz.  The ID numbers given in the MWA tables (154 and 185 MHz and light curve data) are common to those tables i.e. when the same ID number appears at both frequencies, this indicates a source that is cross-matched at the two frequencies.  The ID numbers given for the GMRT data in Table 3 are specific to the GMRT data - they are not the same as the ID numbers provided for the MWA data.

\begin{table*}[ht]
\centering
\begin{tabular}{ccccccc}
\hline
RA (J2000) & $\Delta$RA (J2000) & Dec (J2000) & $\Delta$DEC (J2000) & Flux (mJy) & $\Delta$Flux (mJy) & ID \\
\hline
170.3635 & 0.0022 & -8.2628 & 0.0050 & 367 & 9 & 0 \\
170.9044 & 0.0022 & -8.1030 & 0.0041 & 311 & 3 & 1 \\
171.9504 & 0.0044 & -8.0235 & 0.0050 & 452 & 45 & 2 \\
173.6057 & 0.0027 & -8.0053 & 0.0062 & 4104 & 74 & 3 \\
175.7092 & 0.0050 & -7.8707 & 0.0087 & 425 & 28 & 4 \\
174.0945 & 0.0037 & -7.8521 & 0.0087 & 353 & 21 & 5 \\ \hline
\end{tabular}
\caption{Example catalog from the MWA at 154~MHz. ID is the source identification, used in place of a position based name. A full version of this table and the corresponding table at 184~MHz is available as a Machine Readable Table (in CSV format).}
\label{tab:cat_154}
\end{table*}

\begin{table*}[ht]
\centering
\small
\begin{tabular}{ccccccccccccc}
\hline
 RA & $\Delta$RA             & Dec & $\Delta$Dec            & $S$  & $\Delta S$  & a  & err a                 & b & err b & pa & err pa & ID \\
 \multicolumn{2}{c}{J(2000)} &  \multicolumn{2}{c}{J(2000)} & Jy         &    Jy               & \multicolumn{2}{c}{arcsec} & \multicolumn{2}{c}{arcsec} & \multicolumn{2}{c}{deg} & \\
 \hline
171.9716 & 0.0001 & -8.5370 & 0.0001 & 0.153 & 0.004 & 43.45 & 0.02 & 32.69 & 0.04 &   4.6 & 0.1 & 1 \\
172.0113 & 0.0004 & -8.5207 & 0.0005 & 0.037 & 0.005 & 24.24 & 0.04 & 21.72 & 0.05 &   1.4 & 1.2 & 2 \\
172.4271 & 0.0002 & -8.5068 & 0.0002 & 0.078 & 0.005 & 25.93 & 0.00 & 24.41 & 0.00 &   0.0 & 0.9 & 3 \\
171.0500 & 0.0003 & -8.5007 & 0.0002 & 0.076 & 0.004 & 54.17 & 0.05 & 37.61 & 0.03 &  87.7 & 0.1 & 4 \\
172.8025 & 0.0002 & -8.4320 & 0.0003 & 0.071 & 0.005 & 27.95 & 0.03 & 25.42 & 0.03 &  -2.3 & 0.6 & 5 \\
170.9648 & 0.0006 & -8.4236 & 0.0006 & 0.031 & 0.005 & 21.00 & -1   & 21.00 & -1   &  89.7 & -1  & 6 \\ \hline
\end{tabular}
\caption{Example catalog from the TGSS image. ID is the source identification, used in place of a position based name. A full version of this table is available as a Machine Readable Table (in CSV format).  When the source finding algorithm cannot assign a meaningful error on the morphology parameters (in the low signal to noise detections), $-1$ is entered into the relevant table entry (see ID 6 in the above table snippet).}
\label{tab:cat_TGSS}
\end{table*}

The final set of MWA images, after source finding, yields a total of 1,085 radio sources at 154 MHz and 1,468 at 185 MHz, over 314 square degrees, at angular resolutions of $\sim$4\arcmin.  The GMRT images, after source finding, yields a total of 7,445 radio sources over the same field, at an angular resolution of $\sim$0.3\arcmin.  Thus, the overall survey covers multiple epochs of observation, spans approximately 140 - 200 MHz, is sensitive to structures on angular scales from arcseconds to degrees, and is contemporaneous with the K2 observations of the field over a period of approximately one month.

\subsection{Discussion}

The resulting survey is significant in its own right, in terms of the parameter space it occupies.  However, it is also interesting as the first dedicated radio survey of a K2 field.  Thus, the survey provides a low frequency, multi-resolution, radio component for multi-wavelength investigations of the more than 21,000 K2 targets in this field.

No transient or variable radio sources were detected from the multi-epoch MWA data above a level of 3$\sigma$ using our blind search technique, consistent with similar recent transient surveys of other fields at similar frequencies.  For example, at low frequencies with the MWA 32-tile prototype, \citet{bel13} found no astrophysical transient radio sources at 154 MHz, placing an upper limit on the sky density of $\rho<7.5\times10^{-5}$ deg$^{-1}$ with flux densities $>$5.5 Jy for characteristic timescales of 26 min and 1 year.  Similarly, and more constraining, \citet{row16} undertook a transient survey with the 128-tile full MWA, finding no transient sources greater than 0.285 Jy at 182 MHz on timescales between 28 seconds and 1 year, with $\rho<4.1\times10^{-7}$ deg$^{-1}$ on the shortest timescales.  With the LOFAR telescope, one convincing transient event has been reported recently, by \citet{ste16}.  At current flux density limits, transient radio sources at low radio frequencies appear to be very rare.  Murphy et al. (2016, in preparation) find a low frequency transient rate of $<1.8\times10^{-4}$ per sq. deg. on timescales of 1 to 3 years, from a large-scale comparison of TGSS and MWA datasets.  This limit is an order of magnitude better than previous limits on these timescales.  Long timescale studies probe phenomena that are transient on shorter timescales, in this case ranging from stellar flares to AGN.

Aside from the blind transient and variability search performed here, it is possible to examine our survey data for specific target lists, to search for radio counterparts to targets defined at other wavelengths, or to place limits on the radio emission from such targets.  A significant target class for K2 observations is cool dwarf stars that can produce significant optical and radio flares, in stellar classes M, L, and T.  We thus briefly examine K2 Field 1 target lists in this category, as an example application of our survey data.

Excellent overviews of radio emission from stars can be found in \citet{gud02} and (in particular for flare stars, including at metre wavelengths relevant to our survey) in \citet{bas90}.  Examples of strong radio flares at metre wavelengths, associated with optical flares, from dwarf stars are shown in \citet{spa76}; radio flares at frequencies between 200 and 400 MHz, with durations of hundreds to thousands of seconds and amplitudes of hundreds of mJy, from 13 flares detected during approximately 60 hours of observations.

In the set of K2 Field 1 target lists, a number of approved programs targeted variable optical emission from dwarf stars, including: G01008 (one object); G01010 (seven objects); G01044 (four objects); and G01052 (124 objects).

To illustrate the use of our survey data, we took the objects defined by the target lists above and examined them against our catalogs from the MWA and GMRT observations.  Cross matching the K2 target list with the MWA and GMRT catalogs, we find that one of the K2 targets can be plausibly associated with a radio source in the MWA catalog.  However, given the number of targets and the MWA angular resolution, we expect approximately one false match.  Thus, we attribute this single match to chance rather than an astrophysical association of quiescent emission with a dwarf star.  For completeness, the object in question has a K2 ID of 201548872, appearing in K2 programs GO1011\_SC, GO1010\_LC, GO1060\_LC, and GO1075\_LC (RA=167.541711$^{\circ}$; DEC=1.270302$^{\circ}$; mag=17.866).

Additional MWA and GMRT data exist for other K2 Fields and will be the subject of future publications.  Likewise, the K2 Field 1 observations described in this paper were accompanied by SkyMapper observations (see Figure 2) and a comparison of the MWA, GMRT, and SkyMapper data is deferred to a later publication.

\section{Acknowledgements}
We thank an anonymous referee for comments that significantly improved this paper.  This scientific work makes use of the Murchison Radio-astronomy Observatory, operated by CSIRO. We acknowledge the Wajarri Yamatji people as the traditional owners of the Observatory site.  Support for the operation of the MWA is provided by the Australian Government Department of Industry and Science and Department of Education (National Collaborative Research Infrastructure Strategy: NCRIS), under a contract to Curtin University administered by Astronomy Australia Limited. We acknowledge the iVEC Petabyte Data Store and the Initiative in Innovative Computing and the CUDA Center for Excellence sponsored by NVIDIA at Harvard University. We thank the staff of the GMRT that made these observations possible. GMRT is run by the National Centre for Radio Astrophysics of the Tata Institute of Fundamental Research.  The National Radio Astronomy Observatory is a facility of the National Science Foundation operated under cooperative agreement by Associated Universities, Inc.  This research was conducted by the Australian Research Council Centre of Excellence for All-sky Astrophysics (CAASTRO), through project number CE110001020.  KM acknowledges support from the Hintze Foundation.  PJ acknowledges support via the NRAO Reber Doctoral Fellowship.  This research has made use of NASA's Astrophysics Data System.  This research has made use of the NASA/IPAC Extragalactic Database (NED) which is operated by the Jet Propulsion Laboratory, California Institute of Technology, under contract with the National Aeronautics and Space Administration. 

{\it Facility:} \facility{MWA}, \facility{GMRT}.

\end{document}